\documentstyle[prd,aps,eqsecnum,preprint,tighten]{revtex}
\draft
\begin{document}
\title{ 		Quasi-particle characteristics
               of the one-dimensional polaron       }

\author{G.~Ganbold
\thanks{
E-mail: ganbold@thsun1.jinr.ru;  
Permanent address:
Institute of Physics and Technology, Mongolian Academy of Sciences,
210651 Ulaanbaatar, Mongolia  }}

\address{Bogoliubov Laboratory of Theoretical Physics \\
Joint Institute for Nuclear Research, Dubna, 141980, Russia}

\date{September 15, 1999}

\maketitle

\begin{abstract}
   The main quasi-particle characteristics of the one-dimensional
polaron are estimated within and beyond the most general Gaussian
approximation at arbitrary electron-phonon coupling. We have derived
explicitly the ground-state energy and the effective mass in the
weak- and strong-coupling regimes. For arbitrary coupling, the
Gaussian leading-order term of the polaron self energy improves the
corresponding Feynman estimate and represents the lowest upper bound
available. We have calculated the next non-Gaussian corrections.
Taking into account systematically higher-order corrections does not
perturb considerably the obtained results.
\end{abstract}

\thispagestyle{empty}

\section{Introduction}

A conducting electron moving slowly in a polar crystal interacts
with lattice vibrations and  forms  a quasi-particle,  which is
commonly called a {\sl polaron}. Traditionally, the polaron problem
has been considered in three dimensions \cite{land33}-\cite{devr72}.
In recent years, polaronic effects have been observed in
low-dimensional systems \cite{suyu83}-\cite{lork90} that has
stimulated many theoretical enhancements of the conventional theory
by lowering the spatial dimension \cite{gros76}-\cite{quin94}.

The polaron confined in one dimension ($d=1$) is usually modeled by
the following Hamiltonian \cite{dega90,peet91}
\begin{equation}
H = \frac{p^2}{2}+\sum_{k} a^{\dagger}_{k} a_{k}
+ \sqrt{\frac{\alpha}{L}} \,\sum_{k}
\Big(a^{\dagger}_{k} e^{-i k r} - a_{k} e^{i k r}\Big)\,,
\label{hamilt}
\end{equation}
where ${p}$ and ${r}$ denote the momentum and position of the
electron; ${k}$, $a_{k}$ and $a^{\dagger}_{k}$ are the wave-vector,
annihilation and creation operators of a phonon; $L$ is the length
of lattice crystal and the interaction is characterized by the
dimensionless coupling constant $\alpha\ge 0$. Here we are using
appropriate units, such that, $m_b=\omega_{LO}=\hbar=c=1$, where
$m_b$ is the electron bare mass, $\omega_{LO}$ denotes the constant
frequency of the longitudinal optical phonons.

Various methods have been devoted to investigate the properties
of the system described by (\ref{hamilt}), in particular, the
ground-state energy (GSE) and the effective mass (EM). However,
strict results for these quantities are known only in the limiting
cases of the weak- ($\alpha\to 0$) and the strong-coupling
($\alpha\to\infty$) regime \cite{gros76,lesc89,peet91}.

Among the variety of approaches providing reasonable results for
finite $\alpha$, the Feynman path-integral approach \cite{feyn55}
stands out for its flexibility and all-coupling nature. Following
this formalism one can integrate out the phonon variables in
(\ref{hamilt}) and define the free energy of the polaron
${\cal F}_{\beta}(\alpha)$ as follows:
\begin{equation}
e^{-\beta {\cal F}_{\beta}(\alpha)} = \oint\! \delta r\,
e^{-S_{0}[r] + \alpha W[r]}\,, \qquad S_{0}[r]
\mathrel{\mathop=^{\rm def}}  {1\over 2}\int\limits_{0}^{\beta}
\!dt\, \left( {d r(t) \over dt } \right)^2 \,,
\label{frener}
\end{equation}
where $\beta$ is the inverse temperature, $S_{0}[r]$ represents
the free part of the polaron action while the interaction is
described by
\begin{equation}
\alpha W[r] \mathrel{\mathop=^{\rm def}} {\alpha\over\sqrt{2}}
\int\!\!\!\!\int\limits_{0}^{\beta}\! dt ds \, \frac{e^{-|t-s|}
+e^{-\beta + |t-s|}}{1-e^{-\beta}}\,\delta(|r(t)-r(s)|) \,.
\label{action}
\end{equation}
Functional integration symbol $\oint\! \delta r $ in (\ref{frener})
indicates path integration performed over all closed paths
$r(0)=r(\beta)=x$ followed by ordinary integration over $x$. Hereby,
the free energy of the non-interacting phonons is already subtracted,
so the standard normalization condition ${\cal F}_{\beta}(0)=0$ is
obeyed.

Thus, the original many-body model (\ref{hamilt}) has been
transformed into an effective one-particle problem. However, the
nonlocality and $\delta$-singularity arising in $W[r]$ prevent any
further analytic treatment for finite $\alpha$. For higher dimensions
($d>1$), another Coulomb-like singularity arises in (\ref{action}).
Among the approximate methods, Feynman's all-coupling variational
estimate \cite{feyn55} has been particularly successful (for $d=3$)
to interpolate smoothly between the weak- and the strong-coupling
regime of the ground state ($\beta\to\infty$) by introducing the
following retarded oscillator
\begin{equation}
\alpha W_{F}[{\bf r}] \mathrel{\mathop=^{\rm def}}
- C\!\int\!\!\!\!\int\limits_{0}^{\beta} \! dt ds \,
e^{-\omega\,|t-s|}\,({\bf r}(t)-{\bf r}(s))^2
\label{feynact}
\end{equation}
with adjustable positive parameters $C$ and $\omega$. A combination
of the Jensen-Peierls inequality with the variational principle
leads to the Feynman upper bound to the exact GSE. A similar scheme
has also been suggested by Feynman to estimate the EM. Further
improvement of Feynman's method has been made (for $d=3$) by
increasing the number of trial oscillators \cite{okab71,saya91}.

An essential generalization of the Feynman method has been proposed
in \cite{agl80a} and \cite{sait80} independently. In \cite{agl80a}
the Feynman exponential $C\exp(-\omega\,|t-s|)$ has been changed to
an isotropic trial function $f(t-s)\ge 0$ obeying the periodical
condition $f(t-\beta)=f(t)$. Appropriate variational optimization
has lead to a coupled integral equation for function $f(t)$. By
solving it numerically one obtains the best upper bound available
for $d=3$ \cite{agl80a}. According to \cite{peet91} this result can
be re-scaled into another spatial dimension.

An improvement over the Feynman estimate in a nonvariational way
has been made by applying the Gaussian-equivalent representation
method \cite{efim90,efim95}. According to this method, the polaron
action in arbitrary dimensions $d>1$ can be identically transformed
into an equivalent form
\begin{equation}
S_{GER}[{\bf r}] = {1\over 2} ({\bf r},D^{-1}{\bf r})
+ :\! S_{2}[{\bf r}] \! : \,, \qquad ({\bf r},D^{-1}{\bf r})
\mathrel{\mathop=^{\rm def}} \int\!\!\!\!\int\limits_{0}^{\beta}
\!dt ds \, {\bf r}(t) D^{-1}(t,s) {\bf r}(s)\,,
\label{ger}
\end{equation}
where the Green function of the differential operator $D^{-1}(t,s)$
should be derived from a constraint equation which ensures the
normal-ordered form $:S_{2}[{\bf r}]:$ of the new interaction
functional and the absence of any quadratic path configurations in
it. Within this method the entire Gaussian contribution and the next
non-Gaussian correction to the GSE have been calculated for arbitrary
coupling in two and three dimensions \cite{ganb94,dine95}. The
Gaussian term for the GSE represents an upper bound and in particular,
for $d=3$ we observe exact coincidence with the AGL-S result.
Appropriate estimate  for the EM within and beyond the GER method has
been performed in \cite{ganb98}.

   In the present paper we develop the most general Gaussian
approximation for a wide class of path integrals and apply it to the
specific case of one-dimensional polaron by estimating its main
ground-state characteristics and improving the known results.

\section{Polaron Ground-state Properties}

The main quantities of interest, characterizing the quasi-particle
properties of the polaron are the GSE and EM considered at the
zero-temperature limit $\beta\to\infty$. The GSE of the system is
\begin{equation}
E(\alpha)=\lim\limits_{\beta\to\infty} {\cal F}_{\beta}(\alpha) \,.
\end{equation}

The EM of the polaron may be defined in different ways. To define
the GSE and EM simultaneously we consider the polaron partition
function projected at small fixed momentum $p$ as follows:
\begin{equation}
e^{-\beta {\cal F}_{\beta}(p^2,\alpha)} \!=
{1\over\sqrt{2\pi\beta}} \int \limits_{-\infty}^{\infty}\!\! dx\,
e^{-ipx} \!\int\limits_{r(0)=0}^{r(\beta)=x} \!\!\!\!\delta r\,
e^{-S_{0}[r] + \alpha W[r]}
\: \longrightarrow_{_{\hskip -7mm \beta\to\infty}}
\: e^{-\beta E_{p}(\alpha)} \,.
\label{partfun}
\end{equation}
Normalization in (\ref{partfun}) is chosen so that for large $\beta$
we obtain ${\cal F}_{\beta}(p^2,0)=\beta\,p^2/2$. By changing the
integration variable $r(t)\to q(t)+r(t)$ with $q(t)=xt/\beta$ one
goes to conventional closed paths starting and ending at zero and
rewrites (\ref{partfun}) as follows:
\begin{equation}
e^{-\beta {\cal F}_{\beta}(p^2,\alpha)} \! =
{1\over\sqrt{2\pi\beta}} \int \limits_{-\infty}^{\infty}\!\! dx\,
e^{-ipx-{x^2 / 2\beta}}\, Z_{\beta}(\alpha,x) \,,
\label{purpose}
\end{equation}
\begin{equation}
Z_{\beta}(\alpha,x) = \int\limits_{r(0)=0}^{r(\beta)=0}
\!\!\!\!\delta r\, e^{-S_{0}[r]+\alpha W[r+q]}
]\mathrel{\mathop=^{\rm def}}
\int\! d\sigma_{0}[r] \, e^{\alpha W[r+q]} \,,
\qquad \int\! d\sigma_{0}[r] \cdot 1 = 1  \,.
\label{problem}
\end{equation}

Note, going to the Fourier transform for $\delta$-function we can
rewrite the polaron self-interaction in (\ref{action}) as follows:
\begin{equation}
W[r] =  {1\over 2\sqrt{2}\,\pi} \int\!\!\!\!\int
\limits_{0}^{\beta}\! dt ds \,e^{-|t-s|} \!\int\limits_{-\infty}
^{\infty} \!\!dk \, e^{ik(r(t)-r(s))} \mathrel{\mathop=^{\rm def}}
\int\!\! d\Omega_o(t,s,k)\, e^{i k R(t,s)} \,.
\label{interact}
\end{equation}

Since the polaron action is translationally invariant, $E_{p}(\alpha)$
is a continuous function of $p$ and one can expand it around small
momentum as follows:
$E_{p}(\alpha)=E(\alpha)+{p^2 / 2 m^*(\alpha)} + O(p^4)$. Therefore,
the GSE and EM of the polaron are
\begin{eqnarray}
E(\alpha) \!\!\!&=&\!\!\! -\lim\limits_{\beta\to\infty}
{\cal F}_{\beta}(0,\alpha)\,,               \nonumber   \\
m^{*}(\alpha) \!\!\!&=&\!\!\! \left( \left. \lim_{\beta\to\infty}
{\partial^2 \over \partial p^2} {\cal F}_{\beta}(p^2,\alpha)
\right\vert_{_{p=0}} \right)^{-1}\,.
\label{enermass}
\end{eqnarray}

\section{Generalized Gaussian Approximation}

 As is known, no exact evaluation of (\ref{problem}) is available
yet. Our strategy is to extract exactly the most general Gaussian
contribution out of $Z_{\beta}(\alpha,x)$. Then, the remaining
non-Gaussian correction may be systematically estimated to improve
the main approximation and to control the accuracy.

{\bf i.} First, we demonstrate the basis of our method by evaluating
$Z_{\beta}(\alpha,0)$. This simplification does not harm the GSE and
the necessary extension to $x\neq 0$ required for the EM will be
given later. In principle, the interaction functional $W[r]$ may be
more general than that given in (\ref{action}) and (\ref{interact}).
So, the aim is to find an optimal representation for
\begin{equation}
Z_{\beta}(\alpha) = \langle e^{\alpha W[r]} \rangle_{0} \,, \qquad
\langle \bullet \rangle_{0} \mathrel{\mathop=^{\rm def}}
\int\!\! d\sigma_{0}[r] \, \bullet \,, \qquad
\langle \, 1 \, \rangle_{0} = 1 \,.
\label{problem1}
\end{equation}
For $\alpha\ll 1$, the Gaussian contribution in (\ref{problem1}) is
mainly represented by the functional measure $d\sigma_{0}[r]$ and
the influence of the quadratic part entering $W[r]$ is insignificant.
However, as $\alpha$ increases, the nontrivial Gaussian part of $W[r]$
plays considerable role and this drastically diminishes the efficiency
of the original representation (\ref{problem1}). Besides, the first
cumulant $\langle W[r] \rangle_{o}$ is nonzero.

To describe the system more efficiently at finite and large $\alpha$
we go to another functional representation based on the most general
Gaussian measure $d\sigma [r]$ as follows
\begin{equation}
\int\limits_{r(0)=0}^{r(\beta)=0}\!\!\!\!\delta r \,
e^{-{1\over 2} (r,D^{-1}r)} \,\bullet \mathrel{\mathop=^{\rm def}}
\int\!\! d\sigma [r]\, \bullet = \langle \bullet \rangle
\,, \qquad
\langle 1 \rangle = 1 \,,
\label{measur2}
\end{equation}
A specific restriction imposed on the Green function $D(t,s)$ of the
operator $D^{-1}(t,s)$ will be given later. Obviously, for
$\beta\to\infty$ we have
\begin{equation}
\langle r(t)\, r(s) \rangle = D(t-s) \,, \qquad
\langle e^{i k R(t,s)} \rangle = e^{-k^2 F(|t-s|)} \,, \qquad
F(t) \mathrel{\mathop=^{\rm def}}  D(0)-D(t) \,.
\end{equation}

Functional averaging schemes (\ref{problem1}) and (\ref{measur2}) are
related to each other as follows:
\begin{equation}
\langle  e^{T[r]}\, \bullet \rangle
= \langle e^{T[r]}\rangle \,\langle \bullet \rangle_{0}
= e^{-C_{0}} \, \langle \bullet \rangle_{0} \,,
\label{relation}
\end{equation}
where
\begin{equation}
T[r] \mathrel{\mathop=^{\rm def}} {1\over 2}
\left( r,(D^{-1}-D_{0}^{-1}) r \right)  \,, \qquad
C_{0} \mathrel{\mathop=^{\rm def}}
-\ln{\sqrt{\frac{\det D_{0}^{-1}}{\det D^{-1}}}} \,.
\end{equation}

Applying (\ref{relation}) to (\ref{problem1}) we isolate the most
general Gaussian part of the initial PI by performing an identical
transformation as follows
\begin{equation}
\langle \, e^{\alpha W[r]}\,\rangle_{0} =
e^{C_{0}} \langle e^{\alpha W[r]} \rangle =
e^{C_{0}+\langle T[r] \rangle + \alpha\langle W[r]
\rangle } \,\, \langle e^{\alpha\overline{W}[r]} \rangle
\mathrel{\mathop=^{\rm def}} e^{-\beta E_o(\alpha)}\,\langle
e^{\alpha\overline{W}[r]}\rangle\,.
\label{ident}
\end{equation}
In particular, for the one-dimensional polaron we have
\begin{equation}
\alpha\,\overline{W}[r] \mathrel{\mathop=^{\rm def}}
\alpha\!\int\!\!d\Omega_{0}(t,s,k)\, \left\{ e^{ikR(t,s)}
- e^{-k^2 F(t-s)} \right\} - \langle T[r] \rangle  \,.
\label{newint}
\end{equation}
Since all Gaussian parts are concentrated, by definition, in the
Gaussian measure, the new interaction (\ref{newint}) should not
contain any quadratic path configurations. This requirement leads
to the following equality
\begin{eqnarray}
0 &=& \langle T[r]\rangle - {\alpha\over 2} \!\int\!\!
d\Omega_{0}(t,s,k)\, e^{-k^2 F(t-s)}\,k^2\, \langle
[R(t,s)]^2 \rangle                      \nonumber \\
  &=&  {1\over 2} \left( (D^{-1}-D_{0}^{-1}),D \right)
- \alpha \!\int\!\! d\Omega_{0}(t,s,k)\, e^{-k^2 F(t-s)}
\,k^2\, F(|t-s|) \,.
\label{constraint}
\end{eqnarray}
Therefore, the adjustable function $F(t)$ playing a key role in the
new representation  should be derived from the constraint equations:
\begin{eqnarray}
F(t)\!\!\! &=& \!\!\!{1\over\pi}\int\limits_{0}
^{\infty}\!\!dk \,[1-\cos(kt)]\, \widetilde{D}(k) \,,   \nonumber\\
\widetilde{D}(k) \!\!\!&=&\!\!\!\left( k^2+{\alpha\over\sqrt{2\pi}}
\!\int \limits_{0}^{\infty}\!\!dt \, e^{-t}\,[1-\cos(kt)]\,
F^{-3/2}(t) \right)^{-1}\,,
\label{consteq}
\end{eqnarray}
where $\widetilde{D}(k)$ is the Fourier transform of $D(t)$.

Exact analytic solutions to (\ref{consteq}) are available in the
weak- and strong-coupling limit:
\begin{equation}
\widetilde{D}(k) = \left\{
\begin{array}{ll}
\left\{ k^2+\alpha \gamma(k) + \alpha^2 \chi(k) \right\}^{-1}
                + O(\alpha^3)\,,\quad & \alpha\to 0 \,,     \\
\{ k^2+v^2 \}^{-1}+ O(1)\,, \qquad\qquad  v={4\alpha^2/\pi} \,,
                 \qquad & \alpha\to\infty   \,,
\end{array}
\right.
\label{asym1}
\end{equation}
where
$$
\gamma(k) \mathrel{\mathop=^{\rm def}} 2\sqrt{2}
\left(\sqrt{1+\sqrt{1+k^2}}-\sqrt{2}\right) \,,
\quad \chi(k) \mathrel{\mathop=^{\rm def}} {k^2\over 2\pi}
\!\int\limits_{0}^{\infty}\!\! dz \,
{(1+z)^3+|1-z|^3-2 z^3-2\over z^{3/2} (1+z) (k^2+(1+z)^2)} \,.
$$
Note, function $\chi(k)$ can be further simplified, but the result
is a long expression consisting of rational and trigonometric
functions.

Substituting (\ref{constraint}) into (\ref{newint}) we rewrite
the new interaction functional as follows
\begin{equation}
\overline{W}_{2}[r] = \int\!\! d\Omega_{0}(t,s,k)\,e^{-k^2 F(t-s)}\,
\left\{ e^{ik[r(t)-r(s)]+k^2 F(t-s)}-1+{k^2 \over 2}[r(t)-r(s)]^2
\right\} \,.
\label{newinter}
\end{equation}
Note again that $\overline{W}_{2}[r]$ does not contain any quadratic
path configurations. Besides, the first cumulant is trivial
$\langle\overline{W}_{2}[r]\rangle =0$.

Finally, we write
\begin{equation}
e^{-\beta E(\alpha)}=
\langle \, e^{\alpha W[r]}\,\rangle_{0} = e^{-\beta E_o(\alpha)}\,
\langle e^{\alpha \overline{W}_{2}[r]} \rangle \,,
\label{gga}
\end{equation}
where $E_o(\alpha)$ is the general Gaussian contribution to the
one-dimensional polaron GSE.

Equations (\ref{ident})-(\ref{gga}) serve as the basis of the new
representation which we call the {\sl generalized Gaussian
approximation}. The remaining non-Gaussian corrections should be
evaluated by considering the following new PI
\begin{equation}
J_\beta(\alpha)=\langle e^{\alpha \overline{W}_{2}[r]} \rangle \,.
\label{correc}
\end{equation}

Note, by applying the Jensen-Peierls inequality to (\ref{correc})
one obtains
\begin{equation}
J_\beta(\alpha)\ge e^{\alpha\langle\overline{W}_{2}[r]\rangle}=1\,.
\end{equation}
Therefore, $E_o(\alpha)$ represents a {\sl upper bound} to the
GSE of the one-dimensional polaron:
\begin{equation}
E_o(\alpha) \ge  E(\alpha) \,.
\end{equation}

{\bf ii.} The described above scheme can be easily generalized to
the case $x\neq 0$ that is necessary to evaluate the EM of the
polaron. Omitting details of the extension we write the final result
as follows
\begin{equation}
Z_\beta(\alpha,x) = e^{-\beta {\cal E}_{G}(x,\alpha,\beta)}
\cdot \langle e^{\alpha\overline{W}_{2}[r,x]} \rangle \,,
\label{final}
\end{equation}
where the leading-order Gaussian approximation is:
\begin{eqnarray}
e^{-\beta {\cal E}_{G}({x},\alpha,\beta)}\!\!\!&=&\!\!\!
\exp\left\{ {\beta\over 2 \pi}\, \!\!\int\limits_{0}^{\infty}
\!\!dk \, \ln(k^2 \widetilde{D}(k)) + {1\over 2}
\left( D^{-1}-D_o^{-1},D \right)             \right. \nonumber \\
\!\!\!&+&\!\!\! \left. \!\alpha\! \int\!\!d\Omega_o (t,s,k)\,
e^{-{k}^2 F(t-s)}\, \,e^{i{kx} (t-s)/\beta }\right\}
\label{ggaus}
\end{eqnarray}
and the non-Gaussian correction is described by
\begin{eqnarray}
\langle e^{\alpha\overline{W}_{2}[r,x]}\rangle &=& \langle
\exp\left\{ \alpha\!\int\!\! d\Omega_{0}(t,s,k)\,
e^{-k^2 F(t-s)+ikx(t-s)/\beta } \right. \nonumber \\
&& \cdot \left. \left[ e^{ik[r(t)-r(s)]+k^2 F(t-s)} - 1
+ {k^2 \over 2} [r(t)-r(s)]^2 \right] \right\}\rangle \,.
\label{nonGaus}
\end{eqnarray}

{\bf iii.} Now we consider the pure Gaussian approximation to
$Z_\beta(\alpha,x)$ as follows
\begin{equation}
Z_\beta^{G}(\alpha,x) = e^{-\beta {\cal E}_{G}(x,\alpha,\beta)} \,.
\label{gauss}
\end{equation}
Substituting (\ref{ggaus}) and (\ref{gauss}) into (\ref{enermass})
we obtain the leading-order Gaussian approximations to the GSE and
EM as follows:
\begin{eqnarray}
E_o(\alpha)\!\!&=&\!\! -\,\frac{1}{2\pi}\!\int\limits_{0}^{\infty}
\!\!dk\left[ \ln\left(k^2{\widetilde D}(k)\right)
-k^2{\widetilde D}(k) + 1 \right] + \frac{\alpha}{\sqrt{2\pi}}\!
\int\limits_{0}^{\infty} \!\!dt\,e^{-t}\, F^{-1/2}(t)\,,\nonumber\\
m^*_o(\alpha) \!\!&=&\!\!
1+\frac{\alpha}{2\sqrt{2\pi}}\!\int\limits_{0}^{\infty}
\!\!dt\,t^2 \,e^{-t}\, F^{-3/2}(t) \,.
\label{gaussian}
\end{eqnarray}

Taking into account (\ref{asym1}) we obtain the following analytic
solutions
\begin{eqnarray}
E_o(\alpha) &=& \left\{
\begin{array}{ll}
-\alpha - (1/4-2/3\pi)\,\alpha^2 - O(\alpha^3)\,,
                                 & \qquad \alpha\to 0   \,,  \\
-\alpha^2/\pi - O(1)\,,          & \qquad \alpha\to\infty \,,
\end{array}
\right.                                            \nonumber \\
m^{*}_o(\alpha) &=& \left\{
\begin{array}{ll}
1 + \alpha /2 + (3/2-4/\pi)\,\alpha^2 + O(\alpha^3)\,,
                                 & \qquad \alpha\to 0    \,,   \\
(16/\pi^2)\,\alpha^4 + O(\alpha^2)\,, & \qquad \alpha\to\infty \,.
\end{array}
\right.
\label{eomoasym}
\end{eqnarray}

{\bf iv.} The Feynman estimate can be reproduced, if one builds
a convex combination of the two known asymptotical solutions
(\ref{asym1}) as follows
\begin{equation}
\widetilde{D}_F(k)= w/k^2+(1-w)/(k^2+v^2) \,.
\label{feynmod}
\end{equation}
Substituting (\ref{feynmod}) into (\ref{gaussian}) and optimizing
the obtained energy with respect to parameters $\{w,v\}$ one is
able to reproduce a Feynman-type upper bound $E_F(\alpha)$ for the
one-dimensional polaron. It is inferior to the general Gaussian
result  for finite $\alpha$, i.e. $E_F(\alpha)>E_{0}(\alpha)$.
Accordingly, $m^*_F(\alpha)$ deviates slightly from $m^*_0(\alpha)$.
Note, however, that (\ref{feynmod}) is {\it not} the exact solution
to (\ref{consteq}).

\section{Non-Gaussian Correction}

   The Gaussian leading-order terms $E_o(\alpha)$ and $m_o^*(\alpha)$
approximate well the exact GSE and EM of the polaron. To estimate the
influence of the non-Gaussian correction we evaluate (\ref{nonGaus})
by using the following expansion:
\begin{equation}
J_\beta(\alpha,x) = \langle e^{\alpha\overline{W}_{2}[r,x]} \rangle
= \exp\left\{ \alpha\,\langle \overline{W}_{2}[r,x] \rangle
+{\alpha^2\over 2} \left[ \langle (\overline{W}_{2}[r,x])^2\rangle
-\langle\overline{W}_{2}[r,x]\rangle^2\right]+\ldots \right\}  \,.
\label{expan}
\end{equation}
We stress that this is not a conventional perturbation series on the
coupling constant $\alpha$, each term of the exponent in the r.h.s.
of (\ref{expan}) contains $\alpha$ in more complicated way and the
result is a rapidly converging series even for large $\alpha$.

In doing so we restrict ourselves by estimating only up to the second
cumulant in (\ref{expan}). Appropriate analysis performed in the
weak- and strong-coupling regimes for the third cumulant indicates
that taking into account higher-order cumulants results  in only tiny
improvement over the obtained estimate (see, e.g. (\ref{emweak}) and
(\ref{ener3})). We suppose that this picture remains generally valid
in the intermediate region of $\alpha$. We obtain
\begin{equation}
J_\beta(\alpha,x) = \exp \left\{ -\beta\Delta E_2(\alpha)
-\frac{x^2}{2\beta}\Delta M_2(\alpha)- \ldots \right\} \,,
\label{jbeta}
\end{equation}
where
\begin{eqnarray}
\Delta E_2(\alpha) \!\!\!&=&\!\!\!
-\lim\limits_{\beta\to\infty}\frac{\alpha^2}{4\beta}
\int\!\!\!\!\int\limits_{0}^{\beta}\!\!dt\,ds
\!\!\int\!\!\!\!\int\limits_{0}^{\beta}\!\!dx\,dy
\!\!\int\!\!\!\!\!\!\int\limits_{-\infty}^{\infty}\!\!dk\,dp
\,\rho (t,s,x,y) \,, \nonumber \\
\Delta M_2(\alpha) \!\!\!&=&\!\!\!
\lim\limits_{\beta\to\infty}\frac{\alpha^2}{4\beta}
\int\!\!\!\!\int\limits_{0}^{\beta}\!\!dt\,ds
\!\!\int\!\!\!\!\int\limits_{0}^{\beta}\!\!dx\,dy
\!\!\int\!\!\!\!\!\!\int\limits_{-\infty}^{\infty}\!\!dk\,dp
\,\rho (t,s,x,y)\,(k |t-s| + p |x-y|)^2      \,.
\label{secord}
\end{eqnarray}
Above, we have introduced the following correlation functions:
\begin{eqnarray}
\Xi (t,s,x,y)  \!\!\!\!&& \mathrel{\mathop=^{\rm def}}
 F(t-x)+F(s-y)-F(s-x)-F(t-y)
                                            \,, \nonumber \\
\rho (t,s,x,y) \!\!\!\!&& \mathrel{\mathop=^{\rm def}}
 e^{-|t-s|-|x-y|}\,
\exp\left\{-k^2 F(t-s)-p^2 F(x-y)+kp\,\Xi (t,s,x,y) \right\} \,.
\end{eqnarray}

The second-order non-Gaussian contributions in (\ref{secord}) may
be derived analytically for the weak- and strong-coupling limit.
For finite $\alpha$ we integrate out (\ref{secord}) explicitly over
variables $k,p,x$ and $y$ and the remaining double integrals are
calculated numerically.

Finally, taking into account both the leading-order Gaussian and
the second-order non-Gaussian contribution we estimate the
one-dimensional polaron GSE and EM as follows:
\begin{eqnarray}
E_2(\alpha) &=& E_o(\alpha) + \Delta E_2(\alpha) \,,  \nonumber \\
m^*_2(\alpha) &=& m_o^*(\alpha) + \Delta M_2(\alpha) \,.
\label{e59}
\end{eqnarray}

\section{Exact and Numerical Results}

We have calculated analytically the GSE and EM of the one-dimensional
polaron explicitly with accuracy $O(\alpha^4)$ and $O(1)$ in the weak-
and strong-coupling limit, respectively.

\vskip 2mm
{\sl Weak-coupling solutions}
\vskip 2mm

In the weak-coupling limit, various perturbation methods give
convergent series in powers of $\alpha$. For example, analytic
results up to the second order in $\alpha$ read \cite{peet91}
\begin{eqnarray}
E_{PS}(\alpha)   &=& -\alpha
- \left({3/2\sqrt{2}}-1\right)\,\alpha^2-O(\alpha^3)\,,\nonumber \\
m^*_{PS}(\alpha) &=& 1+\alpha/2
+ \left({5 / 8\sqrt{2}}-{1/4}\right)\,\alpha^2+O(\alpha^3) \,.
\label{peetsmonw}
\end{eqnarray}
A Lee-Low-Pines type method taking into account three-phonon
correction to the Davydov phonon coherent state \cite{quin94}
results in the following numerical results
\begin{eqnarray}
E_{CWW}(\alpha) &=& -\alpha - 0.06066\,\alpha^2
- 0.00844\,\alpha^3 + O(\alpha^4)\,, \nonumber \\
m^*_{CWW}(\alpha) &=& 1+\alpha/2 + 0.19194\,\alpha^2
- 0.06912\,\alpha^3 + O(\alpha^4)\,.
\label{cww}
\end{eqnarray}

By using the GGA method we calculate exactly the GSE and EM of the
one-dimensional polaron up to the $\alpha^3$ terms. Knowing
explicitly the weak-coupling solution (\ref{asym1}) we calculate the
second and third-order non-Gaussian corrections to the GSE and EM.
Adding them to the leading-order Gaussian contributions we obtain
\begin{eqnarray}
&& E_0(\alpha) + \Delta E_2(\alpha) + \Delta E_3(\alpha) \nonumber\\
&&~~~~~~~ = -\alpha -(3/\sqrt{8}-1)\,\alpha^2 -(5-63\sqrt{2}/16
+19\sqrt{3}/16-29\sqrt{6}/48)\,\alpha^3 - O(\alpha^4)\,, \nonumber\\
&& m^{*}_0(\alpha) + \Delta m^{*}_2(\alpha)
                   + \Delta m^{*}_3(\alpha)             \nonumber\\
&&~~~~~~~ = 1 + (1/2)\,\alpha  + (5/8\sqrt{2}-1/4)\,\alpha^2
                   + 0.0691096281\,\alpha^3 + O(\alpha^4)  \,.
\label{emweak}
\end{eqnarray}
Note that the remaining non-Gaussian corrections starting from the
fourth order will modify only the neglected term $O(\alpha^4)$.

\vskip 2mm
{\sl Strong-coupling solutions}
\vskip 2mm

According to the Pekar "Produkt-Ansatz" \cite{peka51}, the polaron
ground-state wave function $|\psi\rangle$ in the strong-coupling
regime ($d=3$) is written as a direct product of the electron
$|\Psi\rangle$ and field $|\varphi\rangle$ wave functions, besides,
$|\varphi\rangle$ depends parametrically on $|\Psi\rangle$. Further
development of this method can be found, in particular, in
\cite{bogo49,tyab51}. A reliable numerical computation of the
minimal value of the corresponding Hartree-type functional for
$d=3$ was performed in \cite{miya75}. The most rigorous
investigations for the strong-coupling limit for $d=1$ have been
reported in \cite{gros76,agl80b,lesc89}. Following the Pekar Ansatz
the GSE of the one-dimensional polaron may be found by solving the
following variational task
\begin{equation}
-\lim\limits_{\alpha\to\infty} {E(\alpha)\over\alpha^2}
=\mathop{sup}\limits_{\langle\Psi|\Psi\rangle =1}
\int\limits_{-\infty}^{\infty}\!\! d x \left\{
\Psi(x)\triangle \Psi(x) + 2 |\Psi(x)|^4 \right\} \,.
\label{pekarener}
\end{equation}
The EM is obtained as follows:
\begin{equation}
\lim\limits_{\alpha\to\infty} {m^{*}(\alpha)\over\alpha^4}
={4\over\pi} \!\int\limits_{-\infty}^{\infty}\!\! dk\,k^2\,
|\langle \Psi|e^{ikx}|\Psi\rangle|^2 \,.
\label{pekarmass}
\end{equation}
The corresponding optimized wave function obeying appropriate
boundary conditions is the nontrivial solution of the following
constrained differential equation:
\begin{equation}
\Psi''(x)+4\,\Psi^3(x)-\Psi(x)=0\,, \qquad
\int\limits_{-\infty}^{\infty} \!\!\! dx |\Psi(x)|^2=1\,.
\label{sturm}
\end{equation}
It admits an explicit analytic solution
\begin{equation}
\Psi(x)=\sqrt{2}/(e^{x}+e^{-x})=\left[\sqrt{2}\cosh(x)\right]^{-1}
\label{psi}
\end{equation}
that results in
\begin{eqnarray}
E(\alpha) &=& -{1\over 3}\,\alpha^2 - O(1)\,, \nonumber \\
m^{*}(\alpha) &=& {32\over 15}\,\alpha^4 + O(1) \,.
\label{pekarexact}
\end{eqnarray}
These solutions coincide (after the appropriate re-scaling) with
the exact results reported first in \cite{gros76}.

Our leading-order Gaussian contributions in (\ref{eomoasym}) differ
from the adiabatic ones in (\ref{pekarexact}). The reason is that
as $\alpha$  increases the polaron self-interaction is less well
approximated by a general Gaussian functional. Hence, non-Gaussian
corrections are required to fill this gap. Taking into account
higher-order corrections (\ref{expan}) we obtain a series converging
rapidly to the exact value in (\ref{pekarexact}) as follows:
\begin{eqnarray}
E_o(\alpha) &=& - 0.318310\,\alpha^2 \,, \nonumber \\
E_o(\alpha)+\Delta E_2(\alpha) &=& - 0.327014\,\alpha^2  \,,
                                                    \nonumber \\
E_o(\alpha)+\Delta E_2(\alpha)+\Delta E_3(\alpha)
                &=& - 0.330205\,\alpha^2
\label{ener3}
\end{eqnarray}
Appropriate series for the EM reads
\begin{eqnarray}
m^{*}_o(\alpha) &=&  1.621139\,\alpha^4    \,, \nonumber \\
m^{*}_o(\alpha)+\Delta M_2(\alpha)
            &=& 1.858065\,\alpha^4  \,, \nonumber \\
m^{*}_o(\alpha)+\Delta M_2(\alpha)+\Delta M_3(\alpha)
            &=& 1.966430 \,\alpha^4 \,.
\label{mass2}
\end{eqnarray}

\vskip 2mm
{\sl Intermediate-coupling results}
\vskip 2mm

For finite $\alpha$ we have solved equations (\ref{consteq})
numerically by means of an iteration accepting (\ref{feynmod})
as the first approximation. After the sixth iteration step the
numerical results do not change within the given accuracy. The
obtained intermediate-coupling results for the Gaussian
leading-order contributions and the second-order non-Gaussian
corrections are presented in Tables 1, 2; and depicted in
Figure 1 in comparison with several known data. \\[5mm]

In conclusion, we have evaluated the ground-state energy and
effective mass of the one-dimensional polaron by developing the
generalized Gaussian approximation method. We have obtained exact
analytic solutions to these quantities in the weak-coupling limit
including the $\alpha^3$ term. For the strong coupling, we have
derived the exact GSE and EM by adapting the Pekar adiabatic theory,
then have estimated the same quantities within the GGA method by
developing a systematic iterative scheme to approach rapidly the
exact results.

We have shown that for intermediate coupling the leading-order
Gaussian contribution to the self-energy slightly improves the
Feynman estimate and represents the lowest upper bound available.
Considering the non-Gaussian corrections, we have calculated the
second-order terms for the GSE and EM. Appropriate analysis
performed for the weak- and strong-coupling regimes indicates
that taking into account higher-order corrections results in
tiny improvement over the obtained estimate. We deduct that this
picture changes inconsiderably in the intermediate region of
$\alpha$. Our method works well in the entire range of $\alpha$,
does not require extensive numerical calculations and provides
reliable results rather quickly.

\acknowledgments

The author thanks G.V.~Efimov for useful discussions. I am indebted
to H.~Leschke for reading the manuscript, helpful comments and for
hospitality.



\begin{figure}
\caption{
The ground-state energy and effective mass of the one-dimensional
polaron as functions of the electron-phonon coupling constant
$\alpha$. All curves are normalized to the corresponding Feynman
estimates so that the horizontal dotted lines show the Feynman
approximations. Dashed lines depict the leading-order Gaussian
contributions and solid curves correspond to the corrected results
taking into account the second-order non-Gaussian corrections.  }
\end{figure}

\begin{table}[t]
\begin{center}
\caption{The ground-state energy of the one-dimensional polaron
         in the intermediate-coupling range.}
\vskip 5mm
\begin{tabular}{ccccc}
\hline
\hline
$\alpha$ & $E_{os}$ &   $E_F$   &  $E_o$   &  $E_2$        \\
\hline
\hline
0.5      & -0.5000  & -0.51006  & -0.51027 & -0.51589      \\
1.0      & -1.0000  & -1.04444  & -1.04532 & -1.06710      \\
1.5      & -1.5000  & -1.61314  & -1.61522 & -1.66119      \\
2.0      & -2.0086  & -2.23696  & -2.24044 & -2.31150      \\
2.5      & -2.7043  & -2.95968  & -2.96352 & -3.05052      \\
4.0      & -5.7934  & -6.04779  & -6.04964 & -6.20592      \\
6.0      & -12.155  & -12.4074  & -12.4083 & -12.7345      \\
8.0      & -21.067  & -21.3178  & -21.3183 & -21.8871      \\
\hline
\hline
\end{tabular}
\end{center}
\end{table}

\begin{table}[ht]
\begin{center}
\caption{The effective mass of the one-dimensional polaron
         in the intermediate-coupling range.}
\vskip 5mm
\begin{tabular}{ccccc}
\hline
\hline
$\alpha$ & $m^*_{os}$ & $m^*_F$ &  $m^*_o$  &  $m^*_2$       \\
\hline
\hline
0.5      & 1.25000  &  1.32084  &  1.32246  &  1.31091       \\
1.0      & 1.50000  &  1.88895  &  1.89862  &  1.83258       \\
1.5      & 1.75000  &  3.11732  &  3.15094  &  2.91361       \\
2.0      & 5.85933  &  6.83836  &  6.90814  &  6.13458       \\
2.5      & 31.6793  &  21.4723  &  21.4372  &  19.4943       \\
4.0      & 331.719  &  281.622  &  281.474  &  295.800       \\
6.0      & 1911.78  &  1784.90  &  1784.82  &  1973.85       \\
8.0      & 6302.70  &  6068.77  &  6068.73  &  6821.14       \\
\hline
\hline
\end{tabular}
\end{center}
\end{table}

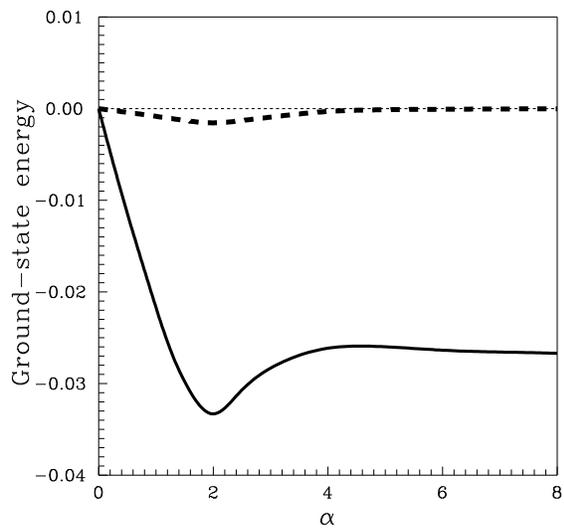
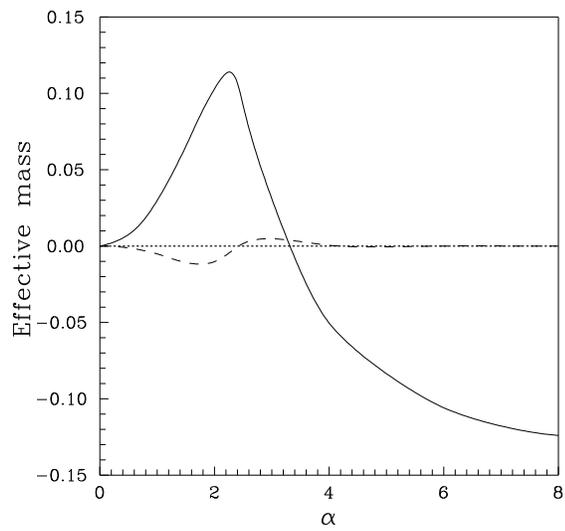
\end{document}